# The Power Inverse Lindley Distribution

Kelly Vanessa Parede Barco
Union Faculty of Campo Mourão, PR, Brazil

Josmar Mazucheli
Statistics Department, Maringá State University, PR, Brazil

Vanderly Janeiro
Statistics Department, Maringá State University, PR, Brazil

## ABSTRACT

Several probability distributions have been proposed in the literature, especially with the aim of obtaining models that are more flexible relative to the behaviors of the density and hazard rate functions. Recently, a new generalization of the Lindley distribution was proposed by Ghitany et al. (2013), called power Lindley distribution. Another generalization was proposed by Sharma et al. (2015), known as inverse Lindley distribution. In this paper, a new distribution is proposed, which is obtained from these two generalizations and named power inverse Lindley distribution. Some properties of this new distribution and study of the behavior of maximum likelihood estimators are presented and discussed. It is also applied considering real data and compared with the fits obtained for already-known distributions. When applied, the power inverse Lindley distribution was found to be a good alternative for modeling survival data.

## 1. Introduction

The Lindley distribution was introduced by Lindley (1958) in the context of Bayes inference. Its density function is obtained by mixing the exponential distribution, with scale parameter $\beta$, and the gamma distribution, with shape parameter 2 and scale parameter $\beta$. Thus, its probability density function is given by:

$$f(x \mid \beta) = p f_1(x \mid \beta) + (1-p) f_2(x \mid 2, \beta),$$

where:

$$p = \frac{\beta}{1+\beta}, \quad f_1(x \mid \beta) = \beta e^{-\beta x} \quad \text{and} \quad f_2(x \mid 2, \beta) = \beta^2 x e^{-\beta x},$$

and therefore:

$$f(x \mid \beta) = \frac{\beta^2}{1+\beta}(1+x)e^{-\beta x}, \tag{1}$$

where $x > 0$ and $\beta > 0$.

The cumulative distribution function is given by:

$$F(x \mid \beta) = 1 - \left(1 + \frac{\beta x}{1+\beta}\right) e^{-\beta x}. \tag{2}$$

Ghitany et al. (2008) studied the properties and applications of this distribution in the context of survival analysis, showing that its mathematical properties are more flexible than those of the exponential distribution.

Since then, many other generalizations of the Lindley distribution have been proposed with the aim of obtaining probability distributions that are more flexible relative to the behaviors of the density and the hazard rate functions.

The inverse Lindley distribution, proposed by Sharma et al. (2015), considers the inverse of a random variable with a Lindley distribution. More specifically, if a random variable $Y$ has a Lindley distribution, then the random variable $X = 1/Y$ follows an inverse Lindley distribution with density and cumulative distribution functions defined, respectively, by:

$$f(x \mid \beta) = \frac{\beta^2}{1+\beta}\left(\frac{1+x}{x^3}\right)e^{-\frac{\beta}{x}}, \tag{3}$$

$$F(x \mid \beta) = \left(1 + \frac{\beta}{1+\beta}\frac{1}{x}\right)e^{-\frac{\beta}{x}}, \tag{4}$$

where $x > 0$ and $\beta > 0$.

Another generalization is the power Lindley distribution, which considers the power of a random variable with a Lindley distribution, proposed by Ghitany et al. (2013). In other words, if a random variable $Y$ has a Lindley distribution, then the random variable $X = Y^{\frac{1}{\alpha}}$ follows a power Lindley distribution.

The power Lindley distribution can be obtained by mixing two distributions, namely the Weibull distribution with shape parameter $\alpha$ and scale parameter $\beta$ and the generalized gamma distribution with shape parameters 2 and $\alpha$ and scale parameter $\beta$. Thus, its probability density function is given by:

$$f(x \mid \alpha, \beta) = pf_1(x \mid \alpha, \beta) + (1-p)f_2(x \mid 2, \alpha, \beta),$$

where:

$$p = \frac{\beta}{1+\beta}, \quad f_1(x \mid, \alpha, \beta) = \alpha\beta x^{\alpha-1}e^{-\beta x^{\alpha}} \quad \text{and} \quad f_2(x \mid 2, \alpha, \beta) = \alpha\beta^2 x^{2\alpha-1}e^{-\beta x^{\alpha}},$$

and therefore:

$$f(x \mid \alpha, \beta) = \frac{\alpha\beta^2}{1+\beta}(1+x^{\alpha})x^{\alpha-1}e^{-\beta x^{\alpha}}, \tag{5}$$

where $x > 0$ and $\alpha, \beta > 0$.

The cumulative distribution function is given by:

$$F(x \mid \alpha, \beta) = 1 - \left(1 + \frac{\beta x^{\alpha}}{1+\beta}\right)e^{-\beta x^{\alpha}}. \tag{6}$$

It can be noticed that the Lindley distribution is a special case of the power Lindley distribution when $\alpha = 1$.

Starting from the inverse Lindley distribution, the power inverse Lindley distribution is introduced in this paper. For that purpose, consider a random variable $Y$ with an inverse Lindley distribution. The random variable $X = Y^{\frac{1}{\alpha}}$ then follows the power inverse Lindley distribution. The probability density function is given by:

$$f(x \mid \alpha, \beta) = \frac{\alpha\beta^2}{1+\beta}\left(\frac{1+x^{\alpha}}{x^{2\alpha+1}}\right)e^{-\frac{\beta}{x^{\alpha}}}, \tag{7}$$

where $x > 0$ and $\alpha, \beta > 0$, $\alpha$ being the location parameter and $\beta$ the scale parameter.

The cumulative distribution and hazard rate functions of the power inverse Lindley distribution are defined, respectively, by:

$$F(x \mid \alpha, \beta) = \left(1 + \frac{\beta}{1+\beta}\frac{1}{x^{\alpha}}\right) e^{-\frac{\beta}{x^{\alpha}}}, \tag{8}$$

$$h(x \mid \alpha, \beta) = \frac{\alpha\beta^2(1 + x^{-\alpha})}{x\big[-\beta + x^{\alpha}(1+\beta)(e^{\frac{\beta}{x^{\alpha}}} - 1)\big]}. \tag{9}$$

It can be noticed that the inverse Lindley distribution is a special case of the power inverse Lindley distribution when $\alpha = 1$.

The aims of this paper are to introduce and to study some properties of the power inverse Lindley distribution. In Section 2 some properties of the density and hazard rate functions are studied. The moments, quantile function, and maximum likelihood estimation are shown in Sections 3, 4, and 5, respectively. In order to study the bias and the mean squared error of the maximum likelihood estimates, some Monte Carlo experiments were conducted, with results shown in Section 6. Section 7 shows the comparison between the new proposed distribution and some other distributions using a real survival data set.

## 2. Behaviors of the Density and Hazard Rate Functions

In Figure 1 the behavior of the probability density function of the power inverse Lindley distribution can be observed for different values of $\alpha$ and $\beta$, showing that the density function (7) is unimodal in $x$.

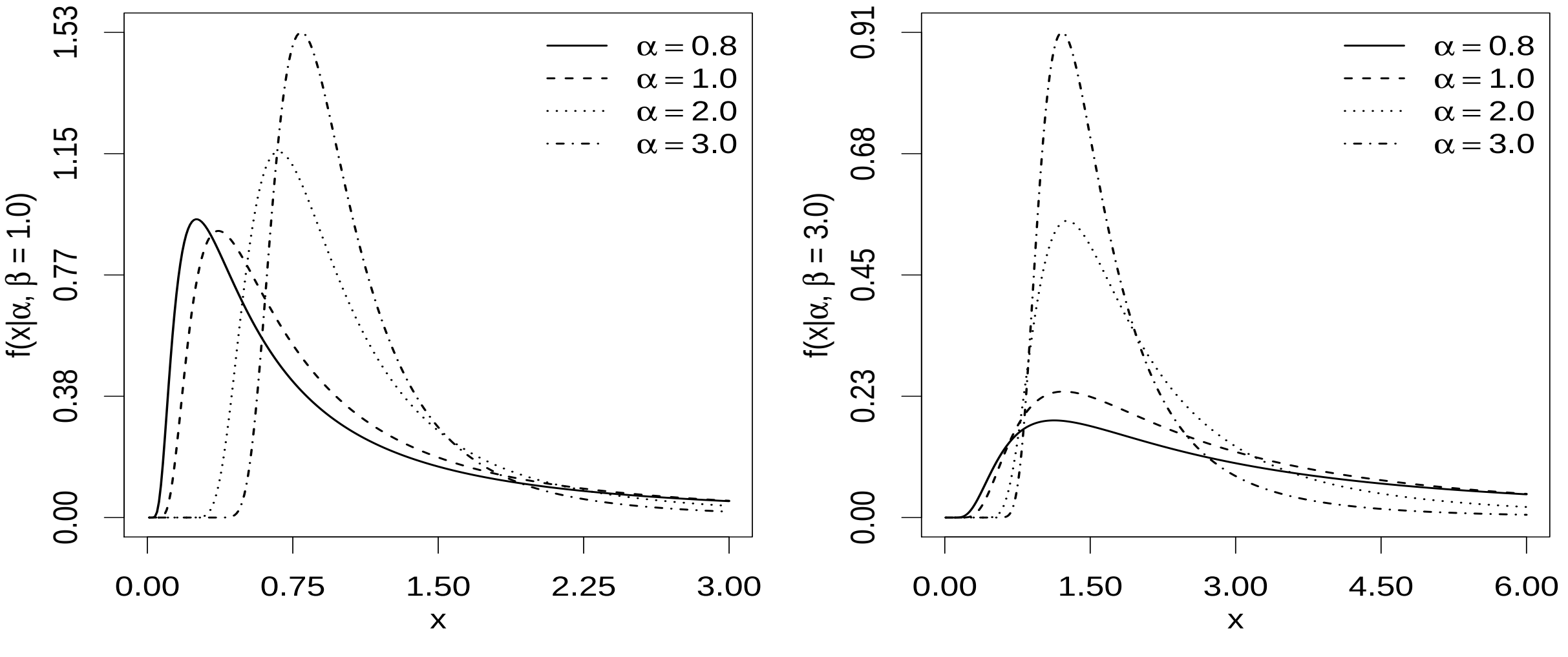

Figure 1: Behavior of the density function of the power inverse Lindley distribution for different values of $\alpha$ and $\beta$.

To determine the mode of the density function (7) of the power inverse Lindley distribution, its maximum must be found. Therefore, the point where the first derivative of the density function is null must be determined.

The first derivative of function (7) is given by:

$$\frac{d}{dx} f(x \mid \alpha, \beta) = \frac{\alpha\beta^2}{1+\beta}\frac{e^{-\frac{\beta}{x^{\alpha}}}}{x^{3\alpha+2}}\psi(x^{\alpha}), \tag{10}$$

where:

$$\psi(x^\alpha) = ax^{2\alpha} + bx^\alpha + c, \quad x^\alpha > 0, \tag{11}$$

with:

$$a = -(\alpha + 1), \quad b = \alpha\beta - 2\alpha - 1, \quad c = \alpha\beta.$$

It is clear that $\psi(x^\alpha)$ is a quadratic function and that $\psi(x^\alpha) = 0$ implies $\frac{d}{dx}f(x \mid \alpha, \beta) = 0$. To determine the mode of the density function, it is then necessary to find the roots of function (11).

In this way, as $a < 0$, function (11) is a parabola that opens downwards and thus has a maximum point. $\psi(0) = c$ and $c > 0$, which means the parabola intersects the y-axis only in positive values, therefore guaranteeing that the maximum point given by $x^\alpha = -\frac{b}{2a}$ is always positive and, consequently, that $b > 0$. In addition, two roots of the function can be determined, one being negative and the other one positive. However, as $x > 0$, only the positive root is of interest, being defined by:

$$x^\alpha = \frac{(\alpha\beta - 2\alpha - 1) + \sqrt{(\alpha\beta - 2\alpha - 1)^2 + 4(\alpha + 1)(\alpha\beta)}}{2(\alpha + 1)}. \tag{12}$$

From (12), the value of $x^\alpha$ that zeroes function (11) is obtained. Consequently, the value of $x$ where (10) is null is also obtained, that is, the mode of the density function of the power inverse Lindley distribution.

Concerning the hazard rate function of the power inverse Lindley distribution, which is shown in Figure 2, it notably has the shape of an upside-down bathtub, therefore being unimodal in $x$.

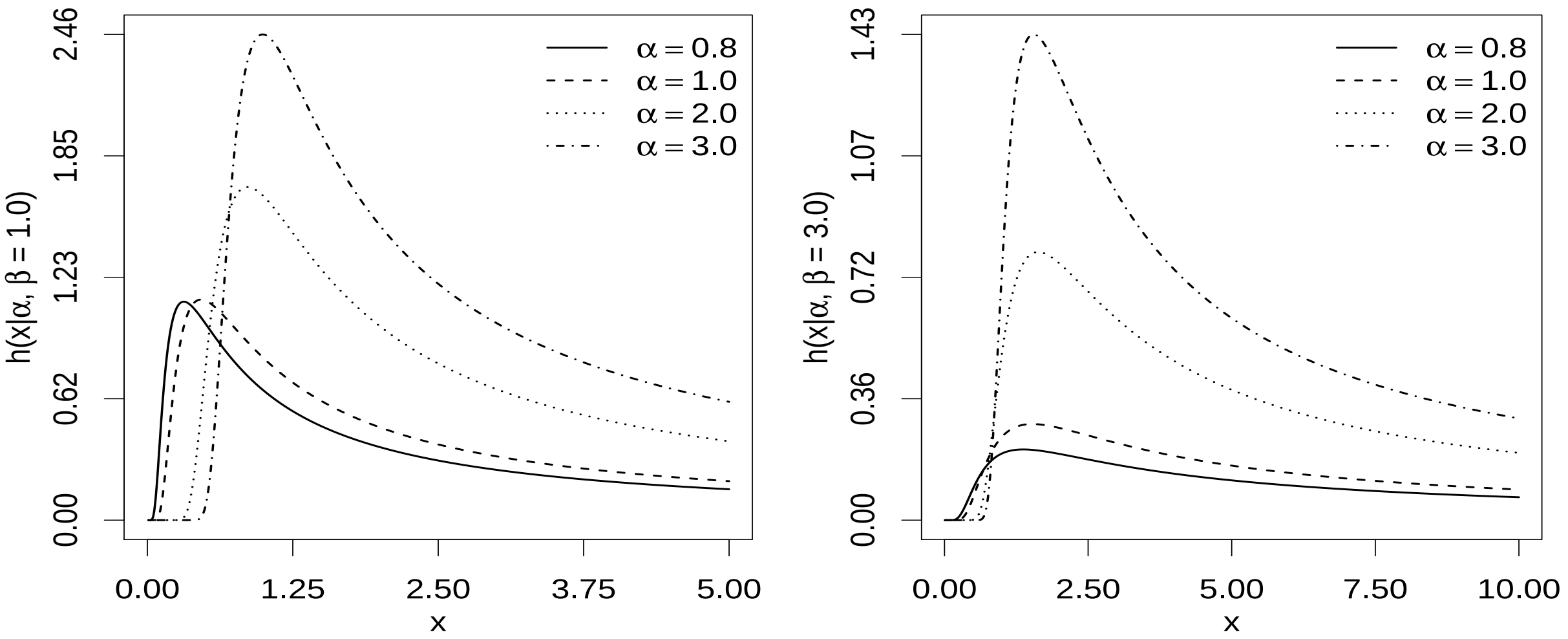

Figure 2: Behavior of the hazard rate function of the power inverse Lindley distribution for different values of $\alpha$ and $\beta$.

As is the case for the density function, the mode of the hazard rate function can also be determined by finding its maximum. The first derivative of the function is given by:

$$\frac{d}{dx}h(x \mid \alpha, \beta) = \frac{\alpha\beta^2 x^{-\alpha}}{x^2[-\beta + x^\alpha(1+\beta)(e^{\frac{\beta}{x^\alpha}} - 1)]^2}\psi(x^\alpha) \tag{13}$$

where:

$$\psi(x^{\alpha}) = ax^{2\alpha} + bx^{\alpha} + c, \quad x^{\alpha} > 0 \tag{14}$$

with:

$$\begin{aligned} a &= (1+\alpha)(1+\beta)(1-e^{\frac{\beta}{x^{\alpha}}}), \\ b &= 1+2(\alpha+\beta+\alpha\beta)+(1+\beta)(\alpha\beta-2\alpha-1)e^{\frac{\beta}{x^{\alpha}}}, \\ c &= \beta(\alpha\beta e^{\frac{\beta}{x^{\alpha}}}+\alpha e^{\frac{\beta}{x^{\alpha}}}+\alpha+1). \end{aligned}$$

It is important to notice that because $e^{\frac{\beta}{x^{\alpha}}} > 1$, $\alpha > 0$ and thus function (14) is a parabola that opens downwards, which has a maximum. Besides, $\psi(0) = c$ and $c > 0$, meaning the parabola intersects the y-axis only at positive values. Analogously to what happens in the density function, the positive solution to equation (14) gives the value of $x^{\alpha}$ of interest and, consequently, the value of $x$ when (13) is null, that is, the mode of the hazard rate function of the power inverse Lindley distribution.

## 3. Moments

For each positive integer $r$, the $r^{th}$ raw moment of the power inverse Lindley distribution can be defined as:

$$E(X^r) = \frac{\left(\alpha\beta^{\frac{r}{\alpha}} - r\beta^{\frac{r}{\alpha}} + \alpha\beta^{\frac{r+\alpha}{\alpha}}\right)\Gamma\left(\frac{\alpha-r}{\alpha}\right)}{\alpha(1+\beta)}. \tag{15}$$

It is worth of note that, for the $r^{th}$ raw moment to exist, the constraint $\alpha > r$ must be satisfied. From (15), the mean and the variance of the power inverse Lindley distribution can be defined, respectively, as:

$$E(X) = \frac{\left(\alpha\beta^{\frac{1}{\alpha}} - \beta^{\frac{1}{\alpha}} + \alpha\beta^{\frac{1+\alpha}{\alpha}}\right)\Gamma\left(\frac{\alpha-1}{\alpha}\right)}{\alpha(1+\beta)}, \tag{16}$$

$$V(X) = \frac{\left(\alpha\beta^{\frac{2}{\alpha}} - 2\beta^{\frac{2}{\alpha}} + \alpha\beta^{\frac{2+\alpha}{\alpha}}\right)\Gamma\left(\frac{\alpha-2}{\alpha}\right)\alpha(1+\beta) - \left(\alpha\beta^{\frac{1}{\alpha}} - \beta^{\frac{1}{\alpha}} + \alpha\beta^{\frac{1+\alpha}{\alpha}}\right)^2\left[\Gamma\left(\frac{\alpha-1}{\alpha}\right)\right]^2}{\alpha^2(1+\beta)^2}. \tag{17}$$

## 4. Quantile Function

In can be noticed in Figure 3 that the cumulative distribution function (8) is strictly increasing and continuous, so the quantile function can be obtained from the equation $Q(u) = F^{-1}(u)$, where $0 < u < 1$.

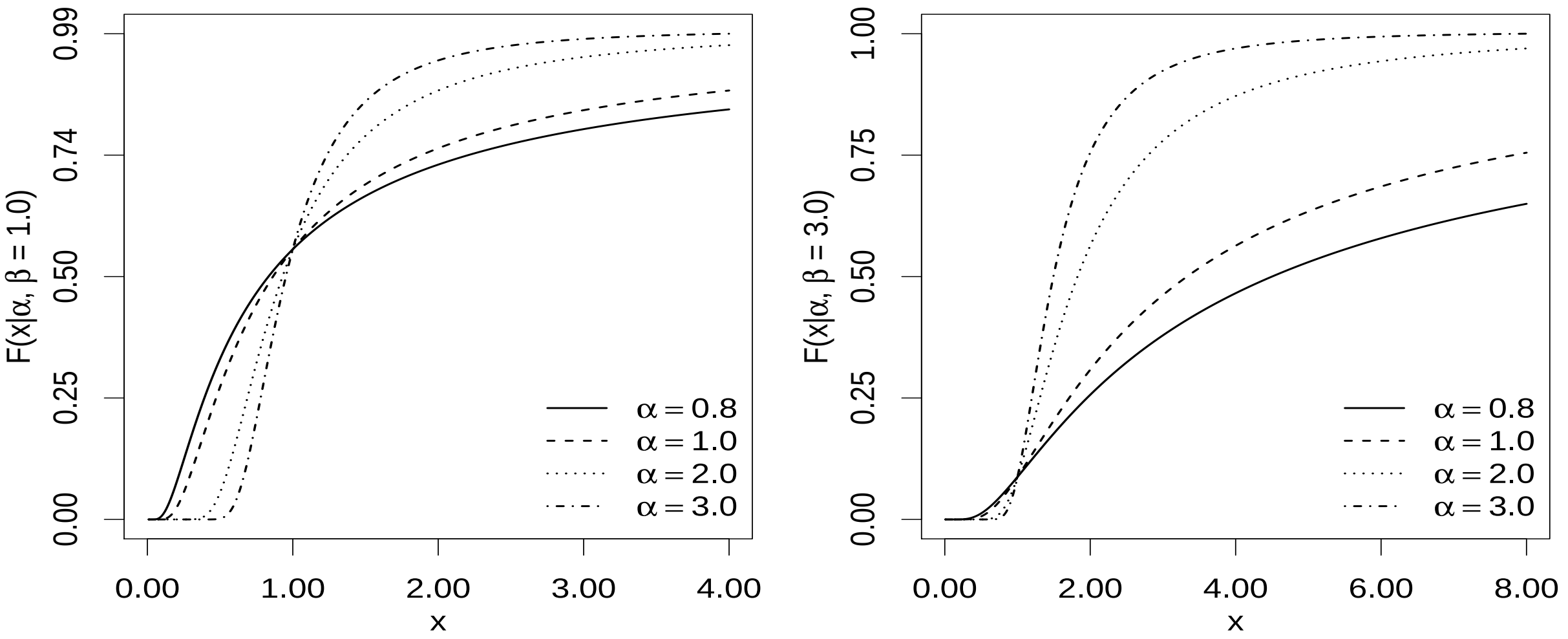

Figure 3: Behavior of the cumulative distribution function of the power inverse Lindley distribution for different values of $\alpha$ and $\beta$.

**Teorem 4.1.** *For any $\alpha, \beta > 0$, the quantile function of the power inverse Lindley distribution is given by:*

$$Q(u) = \left[-1 - \frac{1}{\beta} - \frac{1}{\beta} W_{-1}\left(-u(1+\beta)e^{-(1+\beta)}\right)\right]^{-\frac{1}{\alpha}}, \quad 0 < u < 1,$$

*where $W_{-1}$ denotes the negative branch of the Lambert W function.*

**Proof:** For any fixed $\alpha, \beta > 0$ , let $u \in (1, 0)$. The equation $F(Q(u)) = u$ has to be solved relative to $Q(u)$ for $Q(u) > 0$. Substituting in function (8):

$$\left(1 + \beta + \frac{\beta}{Q(u)^\alpha}\right) e^{-\frac{\beta}{Q(u)^\alpha}} = u(1+\beta). \tag{18}$$

By multiplying both sides of equation (18) by $-e^{-(1+\beta)}$, it becomes:

$$-\left(1 + \beta + \frac{\beta}{Q(u)^\alpha}\right) e^{-(1+\beta+\frac{\beta}{Q(u)^\alpha})} = -u(1+\beta)e^{-(1+\beta)}. \tag{19}$$

To solve equation (19) the Lambert W function, presented by Jodra (2010), must be used. It is a multivalued complex function defined as the solution to the equation $W(z)e^{W(z)} = z$, where $z$ is a complex number.

From equation (19) it can be noticed that $-\left(1 + \beta + \frac{\beta}{Q(u)^\alpha}\right)$ is a Lambert W function of real argument $-u(1+\beta)e^{-(1+\beta)}$. That is:

$$W\left(-u(1+\beta)e^{-(1+\beta)}\right) = -\left(1 + \beta + \frac{\beta}{Q(u)^\alpha}\right). \tag{20}$$

Furthermore, for any $\alpha, \beta > 0$ and $x > 0$, it is immediate that $1 + \beta + \frac{\beta}{Q(u)^\alpha} > 1$ and it can also be verified that $-u(1+\beta)e^{-(1+\beta)} \in (-\frac{1}{e}, 0)$ since $0 < u < 1$. Therefore, by taking into account the properties of the negative branch of the Lambert W function, equation

(20) becomes:

$$W_{-1}\left(-u(1+\beta)e^{-(1+\beta)}\right) = -\left(1+\beta+\frac{\beta}{Q(u)^{\alpha}}\right), \tag{21}$$

which, in turn, implies:

$$Q(u) = \left[-1-\frac{1}{\beta}-\frac{1}{\beta}W_{-1}\left(-u(1+\beta)e^{-(1+\beta)}\right)\right]^{-\frac{1}{\alpha}}. \tag{22}$$

■

From equation (22), the quartiles of the power inverse Lindley distribution can be determined. To that end, the assumption $u = \frac{1}{4}, \frac{1}{2}$ and $\frac{3}{4}$ is made. The quartiles of the power inverse Lindley distribution are then given by:

$$Q\left(\frac{1}{4}\right) = \left[-1-\frac{1}{\beta}-\frac{1}{\beta}W_{-1}\left(-\frac{1}{4}(1+\beta)e^{-(1+\beta)}\right)\right]^{-\frac{1}{\alpha}},$$

$$Q\left(\frac{1}{2}\right) = \left[-1-\frac{1}{\beta}-\frac{1}{\beta}W_{-1}\left(-\frac{1}{2}(1+\beta)e^{-(1+\beta)}\right)\right]^{-\frac{1}{\alpha}},$$

$$Q\left(\frac{3}{4}\right) = \left[-1-\frac{1}{\beta}-\frac{1}{\beta}W_{-1}\left(-\frac{3}{4}(1+\beta)e^{-(1+\beta)}\right)\right]^{-\frac{1}{\alpha}}.$$

## 5. Maximum Likelihood Estimation

Let $x_1, ..., x_n$ be a random sample of size $n$ from the power inverse Lindley distribution. The log-likelihood function is then given by:

$$\begin{aligned} l(\alpha,\beta \mid \mathbf{x}) &= n[\log\alpha + 2\log\beta - \log(1+\beta)] + \sum_{i=1}^{n}\log(1+x_i^{\alpha}) \\ &\quad -(2\alpha+1)\sum_{i=1}^{n}\log x_i - \beta\sum_{i=1}^{n}x_i^{-\alpha}. \end{aligned} \tag{23}$$

Thus, the maximum likelihood estimates $\widehat{\alpha}, \widehat{\beta}$ for $\alpha, \beta$ are the solutions to the non-linear equations:

$$\frac{\partial}{\partial\alpha}l(\alpha,\beta \mid \mathbf{x}) = \frac{n}{\alpha} + \sum_{i=1}^{n}\frac{x_i^{\alpha}\log x_i}{1+x_i^{\alpha}} - 2\sum_{i=1}^{n}\log x_i + \beta\sum_{i=1}^{n}x_i^{-\alpha}\log x_i = 0, \tag{24}$$

$$\frac{\partial}{\partial\beta}l(\alpha,\beta \mid \mathbf{x}) = \frac{n(2+\beta)}{\beta(1+\beta)} - \sum_{i=1}^{n}x_i^{-\alpha} = 0. \tag{25}$$

On the other hand, equation (25) can be rewritten as:

$$\left(\sum_{i=1}^{n}x_i^{-\alpha}\right)\beta^2 + \left(\sum_{i=1}^{n}x_i^{-\alpha} - n\right)\beta - 2n = 0,$$

and therefore, the maximum likelihood estimate $\widehat{\beta}$ is the only solution to the equation above, given by:

$$\widehat{\beta}(\widehat{\alpha}) = \frac{-\left(\sum_{i=1}^{n} x_i^{-\widehat{\alpha}} - n\right) + \sqrt{\left(\sum_{i=1}^{n} x_i^{-\widehat{\alpha}} - n\right)^2 + 8n\sum_{i=1}^{n} x_i^{-\widehat{\alpha}}}}{2\sum_{i=1}^{n} x_i^{-\widehat{\alpha}}}$$

and $\widehat{\alpha}$ is obtained by the solution to the non-linear equation:

$$G(\alpha) = \frac{n}{\alpha} + \sum_{i=1}^{n} \frac{x_i^{\alpha} \log x_i}{1 + x_i^{\alpha}} - 2\sum_{i=1}^{n} \log x_i + \widehat{\beta}(\alpha) \sum_{i=1}^{n} x_i^{-\alpha} \log x_i = 0.$$

To determine confidence intervals for the parameters using the maximum likelihood estimates $\widehat{\alpha}, \widehat{\beta}$ the Fisher information matrix was used, which for a vector of parameters $\boldsymbol{\theta}$ and $n = 1$ is given by:

$$I(\boldsymbol{\theta}) = I_{ij}(\boldsymbol{\theta}) = E\left[-\frac{\partial^2}{\partial\theta_i\partial\theta_j} \log f(X \mid \boldsymbol{\theta})\right].$$

In practice, it is not always possible to compute the integrals involved in the composition of the Fisher information matrix. In this case, the inverse Hessian matrix is used locally in the maximum likelihood estimates.

For the power inverse Lindley distribution, it is possible to determine the Fisher information matrix. Thus, the second-order derivatives of the log-likelihood function are given by:

$$\begin{aligned}
\frac{\partial^2}{\partial\alpha^2} l(\alpha,\beta) &= -\frac{n}{\alpha^2} + \sum_{i=1}^{n} \frac{x_i^{\alpha} \log^2 x_i}{(1 + x_i^{\alpha})^2} - \beta \sum_{i=1}^{n} x_i^{-\alpha} \log^2 x_i, \\
\frac{\partial^2}{\partial\beta^2} l(\alpha,\beta) &= -\frac{n(\beta^2 + 4\beta + 2)}{\beta^2(1+\beta)^2}, \\
\frac{\partial^2}{\partial\alpha\partial\beta} l(\alpha,\beta) &= \sum_{i=1}^{n} x_i^{-\alpha} \log x_i.
\end{aligned}$$

To obtain the Fisher information matrix it is necessary to determine the expectations of the second-order derivatives obtained above taking $n = 1$, without loss of generality. To that end, some integrals have to be calculated. According to Gradshteyn and Ryzhik (2007),

$$\int_0^\infty t^{\nu-1} \log(t) e^{-\beta t} dt = \frac{\Gamma(\nu)}{\beta^\nu}[\psi(\nu) - \log\beta], \quad \nu, \beta > 0.$$

Similarly:

$$\int_0^\infty t^{\nu-1} (\log(t))^2 e^{-\beta t} dt = \frac{\Gamma(\nu)}{\beta^\nu}\left\{[\psi(\nu) - \log\beta]^2 + \zeta(2,\nu)\right\}, \quad \nu, \beta > 0,$$

being $\Gamma(t)$ the gamma function, $\psi(t) = \frac{d}{dt}\log\Gamma(t)$ the digamma function, and $\zeta(z,\nu)$ Riemann's zeta function defined by:

$$\zeta(z,\nu) = \sum_{m=0}^{\infty} \frac{1}{(\nu+m)^z}, \quad z > 1, \quad \nu \neq 0, -1, -2, \ldots$$

From these equations, the required expectations can be calculated to compose the Fisher matrix. In this way, let $T$ be a random sample from the power inverse Lindley distribution. The random variable $X = T^{-\frac{1}{\alpha}}$ then follows the power inverse Lindley distribution, so

$$\begin{aligned}
E\left[\frac{X^\alpha(\log X)^2}{(1+X^\alpha)^2}\right] &= E\left[\frac{T^{-1}}{(1+T^{-1})^2}\left(-\frac{\log T}{\alpha}\right)^2\right] \\
&= \frac{\beta^2}{\alpha^2(1+\beta)}\int_0^\infty \frac{t^{-1}(1+t)(\log t)^2 e^{-\beta t}}{(1+t^{-1})}dt \\
&= \frac{\beta^2}{\alpha^2(1+\beta)}\int_0^\infty \left(1-\frac{1}{1+t}\right)(\log t)^2 e^{-\beta t}dt \\
&= \frac{\beta}{\alpha^2(1+\beta)}\left\{[\psi(1)-\log\beta]^2+\zeta(2,1)-\beta J(\beta)\right\},
\end{aligned}$$

where:

$$J(\beta) = \int_0^\infty \frac{(\log t)^2}{1+t}e^{-\beta t}dt.$$

$$\begin{aligned}
E[X^{-\alpha}(\log X)^2] &= E\left[T\left(-\frac{\log T}{\alpha}\right)^2\right] \\
&= \frac{\beta^2}{\alpha^2(1+\beta)}\int_0^\infty t(\log t)^2(1+t)e^{-\beta t}dt \\
&= \frac{\beta\left\{[\psi(2)-\log\beta]^2+\zeta(2,2)\right\}+2\left\{[\psi(3)-\log\beta]^2+\zeta(2,3)\right\}}{\alpha^2\beta(1+\beta)},
\end{aligned}$$

$$\begin{aligned}
E[X^{-\alpha}\log X] &= E\left[-\frac{T\log T}{\alpha}\right] \\
&= -\frac{\beta^2}{\alpha(1+\beta)}\int_0^\infty t\log t(1+t)e^{-\beta t}dt \\
&= -\frac{\beta[\psi(2)-\log\beta]+2[\psi(3)-\log\beta]}{\alpha\beta(1+\beta)},
\end{aligned}$$

and thus,

$$
\begin{aligned}
I_{11} &= \frac{1}{\alpha^2} - E\left[\frac{X^\alpha(\log X)^2}{(1+X^\alpha)^2}\right] + \beta E\left[X^\alpha(\log X)^2\right] \\
&= \frac{1}{\alpha^2} + \frac{1}{\alpha^2(1+\beta)}\Big\{\beta\left[(\psi(2)-\log\beta)^2+\zeta(2,2)\right] + 2\left[(\psi(3)-\log\beta)^2+\zeta(2,3)\right] \\
&\quad -\beta\left[(\psi(1)-\log\beta)^2+\zeta(2,1)\right] + \beta^2 J(\beta)\Big\}, \\
I_{22} &= \frac{\beta^2+4\beta+2}{\beta^2(1+\beta)^2}, \\
I_{12} &= E[X^{-\alpha}\log X] \\
&= -\frac{\beta[\psi(2)-\log\beta]+2[\psi(3)-\log\beta]}{\alpha\beta(1+\beta)}.
\end{aligned}
$$

To solve these equations the values of the digamma function, locally on 1, 2, and 3, and of Riemann's zeta function, locally on (2,1), (2,2), and (2,3), are necessary and given by:
$\psi(1) = -0.577216, \quad \psi(2) = 0.422784, \quad \psi(3) = 0.922784;$
$\zeta(2,1) = 1.644934, \quad \zeta(2,2) = 0.644934, \quad \zeta(2,3) = 0.394934.$

Under mild regularity conditions, the asymptotic distribution of the maximum likelihood estimator $\widehat{\boldsymbol{\theta}}$ for $\boldsymbol{\theta}$ is given by:

$$\sqrt{n}(\widehat{\boldsymbol{\theta}} - \boldsymbol{\theta}) \to N(0, I^{-1}(\boldsymbol{\theta}))$$

where $I^{-1}(\boldsymbol{\theta})$ is the inverse Fisher information matrix, with:

$$\frac{1}{n}I^{-1}(\boldsymbol{\theta}) = \frac{1}{n}\begin{pmatrix} I_{11} & I_{12} \\ I_{21} & I_{22}\end{pmatrix}^{-1} = \begin{pmatrix} Var(\widehat{\alpha}) & Cov(\widehat{\alpha},\widehat{\beta}) \\ Cov(\widehat{\alpha},\widehat{\beta}) & Var(\widehat{\beta})\end{pmatrix}. \tag{26}$$

The asymptotic $100(1-\delta)\%$ confidence intervals of $\alpha$ and $\beta$, respectively, are then given by:

$$\widehat{\alpha} \pm z_{\frac{\delta}{2}}\sqrt{\widehat{Var(\widehat{\alpha})}}, \quad \text{and} \quad \widehat{\beta} \pm z_{\frac{\delta}{2}}\sqrt{\widehat{Var(\widehat{\beta})}}$$

where $\widehat{Var(\widehat{\alpha})}$ and $\widehat{Var(\widehat{\beta})}$ are the elements of the main diagonal of the matrix defined in (26).

## 6. Simulation Study

This section presents a simulation study with the purpose of verifying the performance of the maximum likelihood estimators of the power inverse Lindley distribution. For this, a simulation algorithm is needed for generating a random sample with density function (7).

By inverting the cumulative distribution function:

$$X_i = \left[-1 - \frac{1}{\beta} - \frac{1}{\beta}W_{-1}\left(-u_i(1+\beta)e^{-(1+\beta)}\right)\right]^{-\frac{1}{\alpha}}, i = 1, \ldots, n, \tag{27}$$

where $u_i$ are values of a uniform random variable in the interval $(0,1)$.

$N = 10000$ pseudo-random samples were generated from the power inverse Lindley distribution using equation (27) with sizes $n = 20, 50, 80, 110, 140, 170, 200$ and parameters $(\alpha, \beta) = (0.8, 0.8), (0.8, 1.0), (0.8, 1.5), (0.8, 2.0), (1.0, 0.8), (1.0, 1.0), (1.0, 1.5), (1.0, 2.0), (1.5, 0.8), (1.5, 1.0), (1.5, 1.5), (1.5, 2.0), (2.0, 0.8), (2.0, 1.0), (2.0, 1.5)$, and $(2.0, 2.0)$.

In addition, the parameters estimates were obtained by the maximum likelihood method. The biases and the mean squared errors of $\widehat{\alpha}$ and $\widehat{\beta}$ were also calculated and are given by:

$$\text{Bias}(\widehat{\theta}) = \frac{1}{N}\sum_{i=1}^{n}(\widehat{\theta}_i - \theta) \quad \text{and} \quad \text{MSE}(\widehat{\theta}) = \frac{1}{N}\sum_{i=1}^{n}(\widehat{\theta}_i - \theta)^2, \quad \text{with} \quad \theta = \alpha, \beta.$$

Figures 4 to 7 present the biases and the mean squared errors of $\widehat{\alpha}$ and $\widehat{\beta}$. In general, the biases and the MSEs tend to zero for large values of $n$, that is, the estimators are asymptotically unbiased.

It is clear in Figure 4 that the bias of $\widehat{\alpha}$ is strictly positive. Moreover, the bias of $\widehat{\alpha}$ increases with increasing $\alpha$ and $\beta$. However, the effects of $\beta$ on the bias of $\widehat{\alpha}$ are very small.

Figure 5 shows that the bias of $\widehat{\beta}$ can be positive or negative. Due to numerical instability for small values of $\beta$, the bias of $\widehat{\beta}$ tends to increase with an increase in sample size and presents an irregular behavior. On the other hand, the bias of $\widehat{\beta}$ always increases with increasing $\beta$.

As for the mean squared errors, Figure 6 shows that the MSE of $\widehat{\alpha}$ increases with an increase in $\alpha$ or $\beta$. However, the effects of $\beta$ in the MSE of $\widehat{\alpha}$ are very small.

Figure 7 shows that the MSE of $\widehat{\beta}$ increases with an increase in $\beta$. However, the effects of $\alpha$ in the MSE of $\widehat{\beta}$ are nearly negligible.

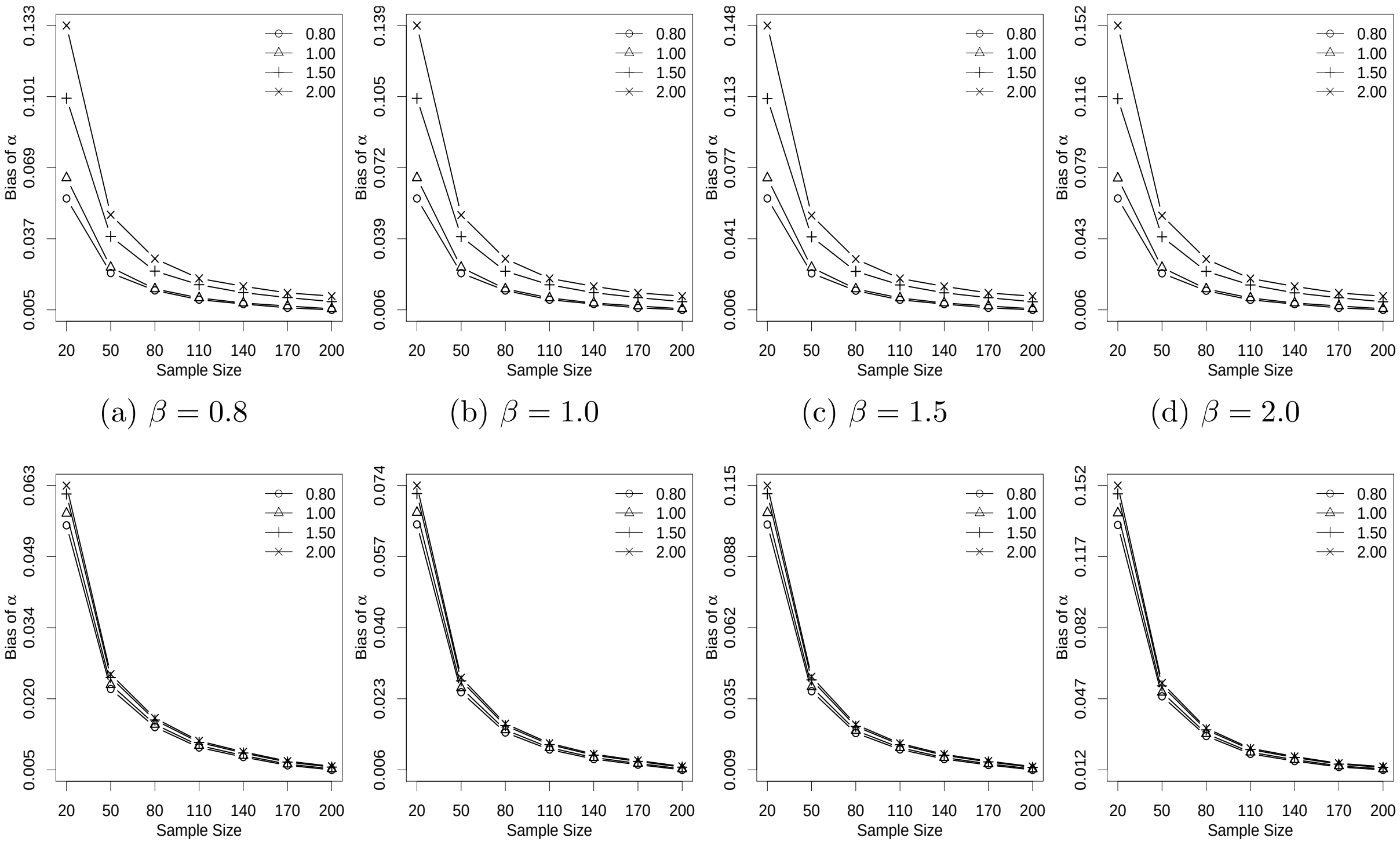

(a) $\beta = 0.8$ (b) $\beta = 1.0$ (c) $\beta = 1.5$ (d) $\beta = 2.0$

(e) $\alpha = 0.8$ (f) $\alpha = 1.0$ (g) $\alpha = 1.5$ (h) $\alpha = 2.0$

Figure 4: Bias of $\widehat{\alpha}$.

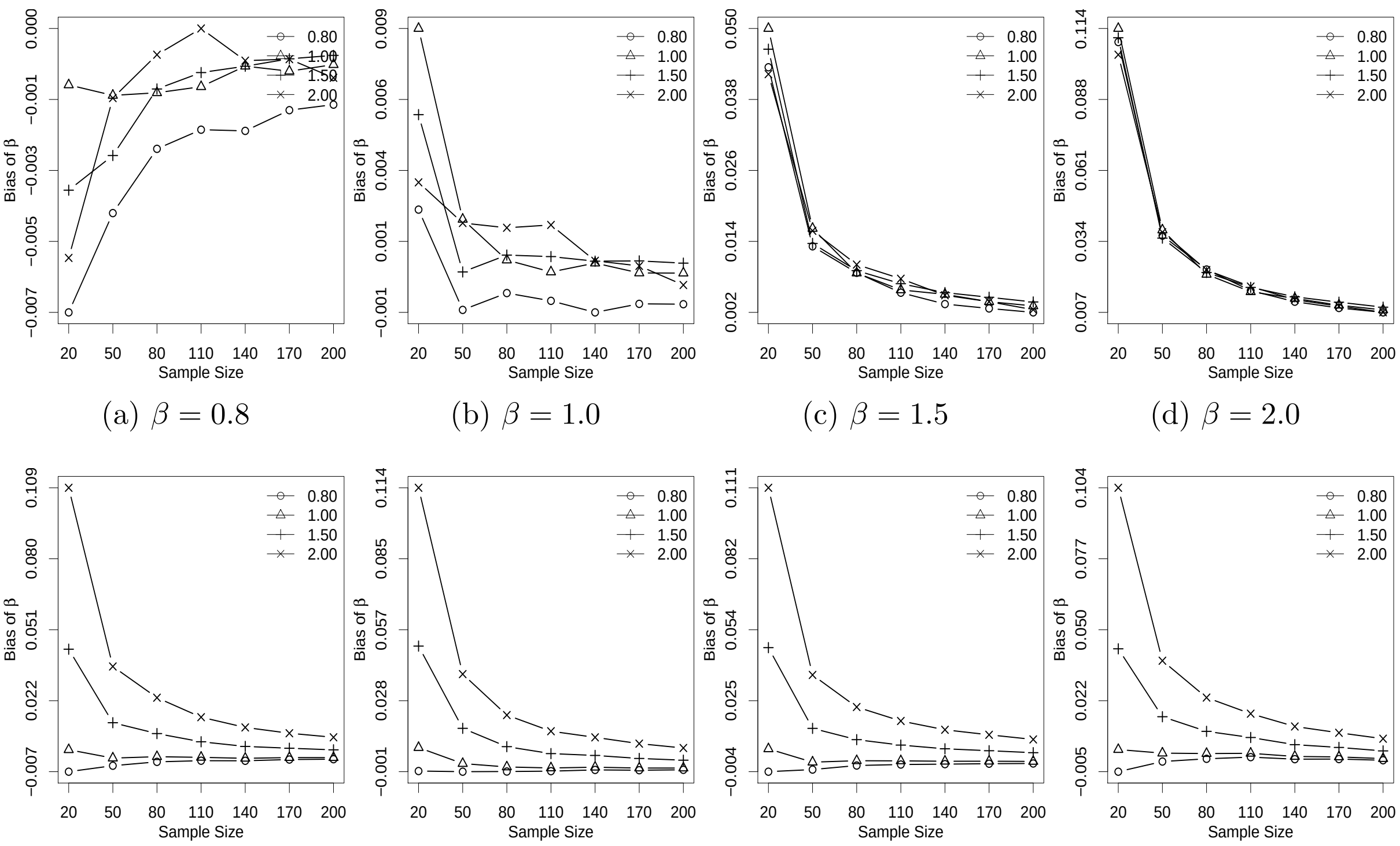

(a) $\beta = 0.8$ (b) $\beta = 1.0$ (c) $\beta = 1.5$ (d) $\beta = 2.0$

(e) $\alpha = 0.8$ (f) $\alpha = 1.0$ (g) $\alpha = 1.5$ (h) $\alpha = 2.0$

Figure 5: Bias of $\widehat{\beta}$.

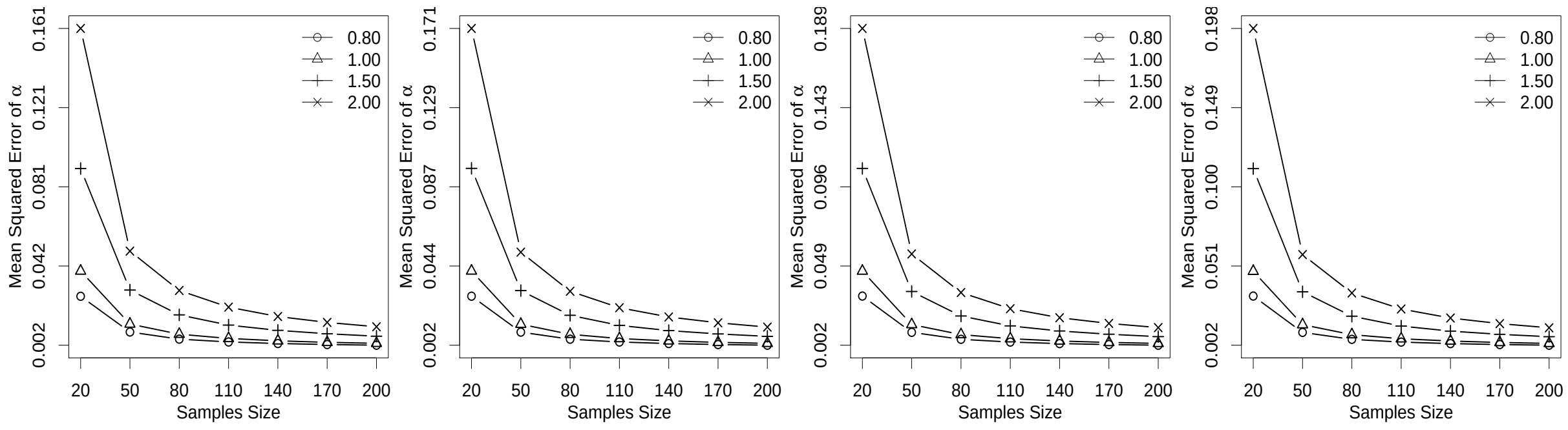

(a) $\beta = 0.8$ (b) $\beta = 1.0$ (c) $\beta = 1.5$ (d) $\beta = 2.0$

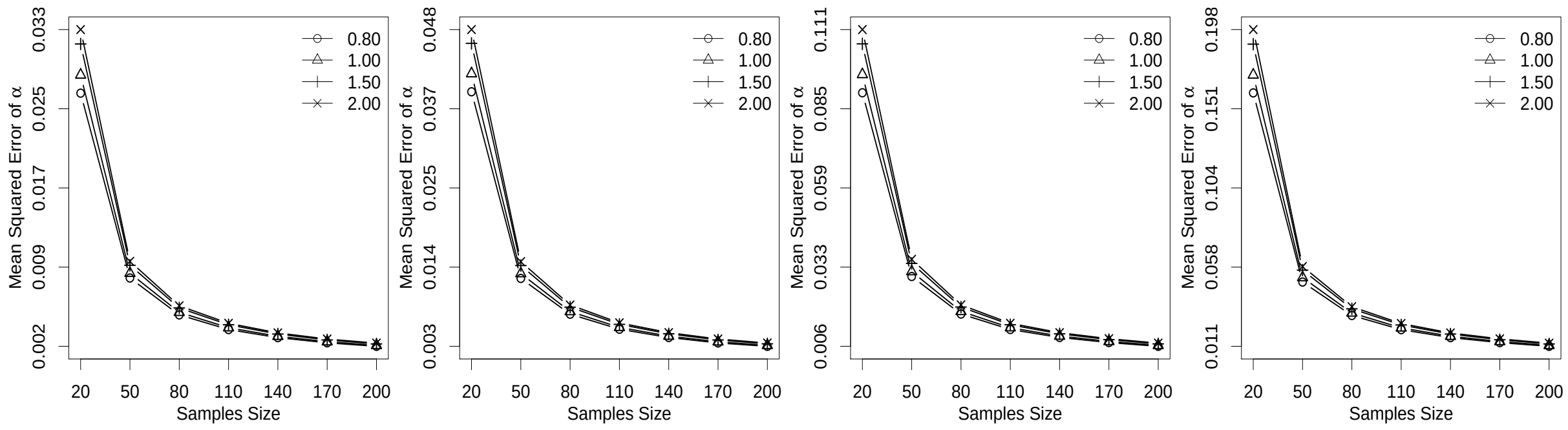

(e) $\alpha = 0.8$ (f) $\alpha = 1.0$ (g) $\alpha = 1.5$ (h) $\alpha = 2.0$

Figure 6: Mean squared error of $\widehat{\alpha}$.

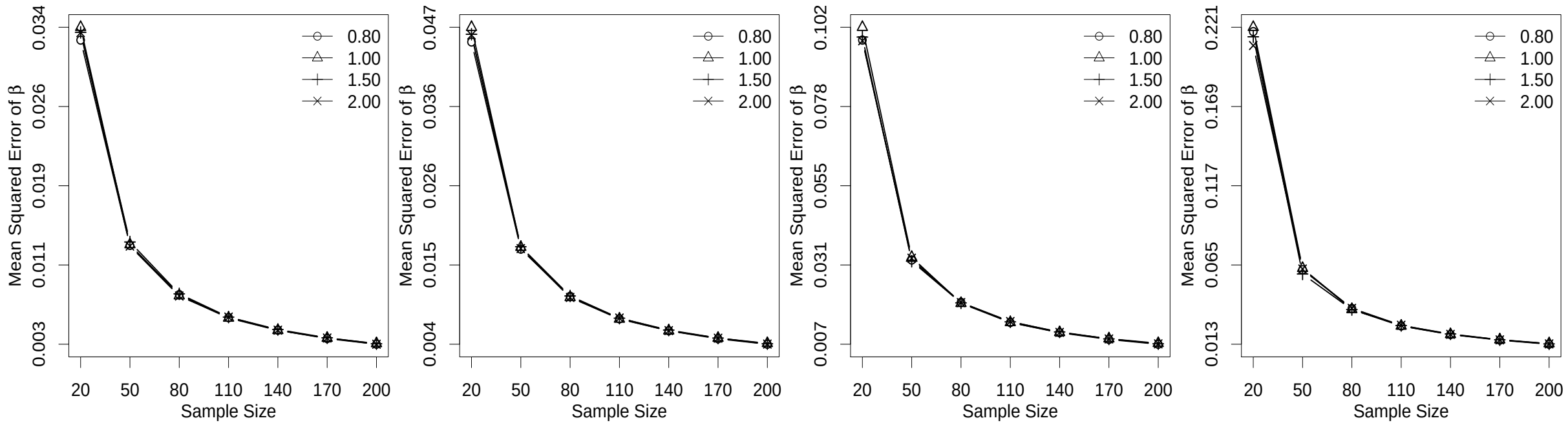

(a) $\beta = 0.8$ (b) $\beta = 1.0$ (c) $\beta = 1.5$ (d) $\beta = 2.0$

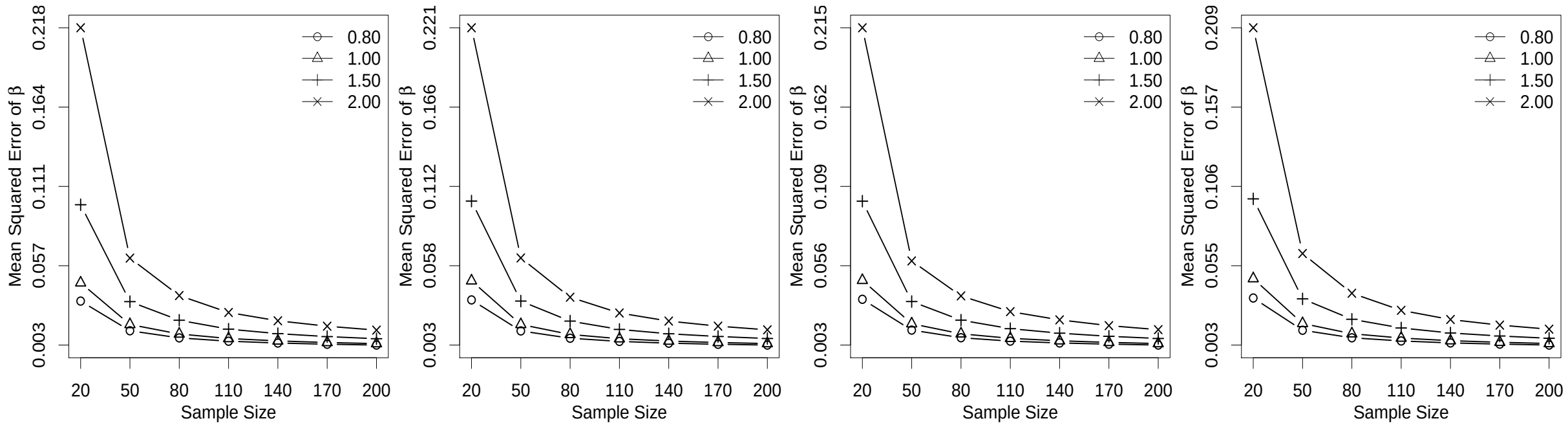

(e) $\alpha = 0.8$ (f) $\alpha = 1.0$ (g) $\alpha = 1.5$ (h) $\alpha = 2.0$

Figure 7: Mean squared error of $\widehat{\beta}$

## 7. Real Data Application

The power inverse Lindley distribution was applied to two sets of data taken from the literature with the objective of evaluating its fit relative to other distributions already present in the literature.

The maximum likelihood method was used to estimate the parameters. For the comparison of the models, the values of $-\log L$, the Akaike information criterion (AIC), and the Bayesian information criterion (BIC), defined respectively by $-2 - \log L + 2q$ and $-2 - \log L + q \log(n)$, where $q$ is the number of estimated parameters and $n$ is the sample size, were taken into account. The most appropriate model corresponds to that which obtains the lowest values for $-\log L$, $AIC$, and $BIC$.

The statistics of the Kolmogorov-Smirnov test (KS), the Anderson-Darling test (AD), and the Cramer-von Mises test (CVM) are also presented, as well as their respective $p$-values. These tests observe the differences between the assumed cumulative distribution function and the empirical cumulative distribution function from the data to verify the fit of the distributions ($p$-value$> 0.05$).

The first data set to be studied was used by Nadarajah et al. (2011) in their work on the generalized Lindley distribution using two parameters. The data shows the relief times of twenty patients receiving an analgesic.

Table 1: Data set 1.

| | | | | | | | | | |
|---|---|---|---|---|---|---|---|---|---|
| 1.1 | 1.4 | 1.3 | 1.7 | 1.9 | 1.8 | 1.6 | 2.2 | 1.7 | 2.7 |
| 4.1 | 1.8 | 1.5 | 1.2 | 1.4 | 3.0 | 1.7 | 2.3 | 1.6 | 2.0 |

Thus, the power inverse Lindley distribution (PIL) is compared with the generalized Lindley (GL), the power Lindley (PL), the inverse Lindley (IL), and the Lindley (L) distributions. The values of $-\log L$, $AIC$, and $BIC$ are presented in Table 3, indicating that the power inverse Lindley distribution (PIL) presents a better fit to the data than the other tested distributions.

Regarding the tests shown in Table 4, it is clear that the power Lindley, the generalized Lindley, and the power inverse Lindley distributions fit the data ($p$-value$> 0.05$), the latter obtaining the lowest test statistics with the largest $p$-values in three tests.

Empirical and theoretical curves of the survival functions of tested distributions are presented in Figure 8. Graphically, it is also possible to observe that the power inverse Lindley distribution was the one which best adjusted to the data set, obtaining a greater approximation between the empirical and the theoretical curves.

The second data set to be studied was used by Mahmoud and Mandouh (2013) in their work on the transmuted Frechet distribution with three parameters. The data set is of simulated strengths of glass fibers and is presented in Table 2.

Table 2: Data set 2.

| | | | | | | | | |
|---|---|---|---|---|---|---|---|---|
| 1.014 | 1.081 | 1.082 | 1.185 | 1.223 | 1.248 | 1.267 | 1.271 | 1.272 |
| 1.275 | 1.276 | 1.278 | 1.286 | 1.288 | 1.292 | 1.304 | 1.306 | 1.355 |
| 1.361 | 1.364 | 1.379 | 1.409 | 1.426 | 1.459 | 1.460 | 1.476 | 1.481 |
| 1.484 | 1.501 | 1.506 | 1.524 | 1.526 | 1.535 | 1.541 | 1.568 | 1.579 |
| 1.581 | 1.591 | 1.593 | 1.602 | 1.666 | 1.670 | 1.684 | 1.691 | 1.704 |
| 1.731 | 1.735 | 1.747 | 1.748 | 1.757 | 1.800 | 1.806 | 1.867 | 1.876 |
| 1.878 | 1.910 | 1.916 | 1.972 | 2.012 | 2.456 | 2.592 | 3.197 | 4.121 |

Similarly, the power inverse Lindley distribution (PIL) is compared with the transmuted Frechet (TF), the power Lindley (PL), the inverse Lindley (IL), and the Lindley (L) distributions. The values of $-\log L$, $AIC$, and $BIC$ are presented in Table 3.

The test results presented in Table 4 indicate that only the transmuted Frechet and the power inverse Lindley distributions fit the data ($p$-value$> 0.05$).

From the results presented for data set 2, it can be observed that the power inverse Lindley distribution obtained the lowest values from the AIC and BIC criteria. This was different for the values of $-\log L$, but with a very subtle difference from the transmuted Frechet distribution. The same is true relative to the goodness-of-fit tests.

In Figure 9 empirical and theoretical curves of the survival functions of the tested distributions are presented.

Graphically, it is possible to observe that the transmuted Frechet distribution provided a good fit to the data, with very little difference relative to that from the power inverse Lindley distribution.

Therefore, it is clear that the power inverse Lindley distribution is a strong competitor to the transmuted Frechet distribution, due to it requiring only two parameters to be estimated.

Table 3: Summary of fitted distributions.

| Data set 1 | | | | | |
|---|---|---|---|---|---|
| Model | MLEs | S.E. | $-\log L$ | AIC | BIC |
| L | $\widehat{\beta} = 0.8161$ | 0.1361 | 30.2496 | 62.4991 | 63.4948 |
| IL | $\widehat{\beta} = 2.2547$ | 0.4089 | 31.7572 | 65.5144 | 66.5101 |
| PL | $\widehat{\alpha} = 2.2529$ | 0.3068 | 20.4320 | 44.8640 | 46.8554 |
| | $\widehat{\beta} = 0.3445$ | 0.0997 | | | |
| GL | $\widehat{\lambda} = 2.5395$ | 0.4492 | 16.4045 | 36.8089 | 38.8004 |
| | $\widehat{\alpha} = 27.8765$ | 19.1608 | | | |
| PIL | $\widehat{\alpha} = 3.9812$ | 0.7041 | 15.4132 | 34.8263 | 36.8178 |
| | $\widehat{\beta} = 6.7190$ | 1.9947 | | | |
| Data set 2 | | | | | |
| Model | MLEs | S.E. | $-\log L$ | AIC | BIC |
| L | $\widehat{\beta} = 0.9383$ | 0.0890 | 85.4760 | 172.9519 | 175.0950 |
| IL | $\widehat{\beta} = 2.0297$ | 0.2053 | 89.3345 | 180.6689 | 182.8121 |
| PL | $\widehat{\alpha} = 2.5761$ | 0.1852 | 45.1260 | 94.2521 | 98.5384 |
| | $\widehat{\beta} = 0.4123$ | 0.0605 | | | |
| TF | $\widehat{\mu} = 4.3142$ | 0.5850 | 19.3693 | 44.7386 | 51.1680 |
| | $\widehat{\sigma} = 1.5491$ | 0.0655 | | | |
| | $\widehat{\lambda} = 0.7778$ | 0.2477 | | | |
| PIL | $\widehat{\alpha} = 5.3904$ | 0.5243 | 20.1115 | 44.2229 | 48.5092 |
| | $\widehat{\beta} = 7.1995$ | 1.2265 | | | |

Table 4: Goodness-of-fit tests.

| Data set 1 | | | | | | |
|---|---|---|---|---|---|---|
| Model | KS | $p$-value | AD | $p$-value | CVM | $p$-value |
| L | 0.3911 | 0.0044 | 3.7504 | 0.0118 | 0.7550 | 0.0086 |
| IL | 0.3695 | 0.0085 | 4.4689 | 0.0053 | 0.9054 | 0.0036 |
| PL | 0.1877 | 0.4815 | 1.0369 | 0.3375 | 0.1754 | 0.3225 |
| GL | 0.1377 | 0.8426 | 0.3344 | 0.9093 | 0.0509 | 0.8764 |
| PIL | 0.1031 | 0.9836 | 0.1560 | 0.9982 | 0.0269 | 0.9873 |
| Data set 2 | | | | | | |
| Model | KS | $p$-value | AD | $p$-value | CVM | $p$-value |
| L | 0.4347 | 2.407e-11 | 15.6600 | 9.524e-06 | 3.2654 | 6.953e-09 |
| IL | 0.4504 | 3.472e-12 | 17.3590 | 9.524e-06 | 3.6203 | 6.812e-10 |
| PL | 0.1909 | 0.0175 | 4.8702 | 0.0033 | 0.8078 | 0.0068 |
| TF | 0.0635 | 0.9472 | 0.4077 | 0.8405 | 0.0530 | 0.8598 |
| PIL | 0.0780 | 0.8091 | 0.5389 | 0.7069 | 0.0714 | 0.7445 |

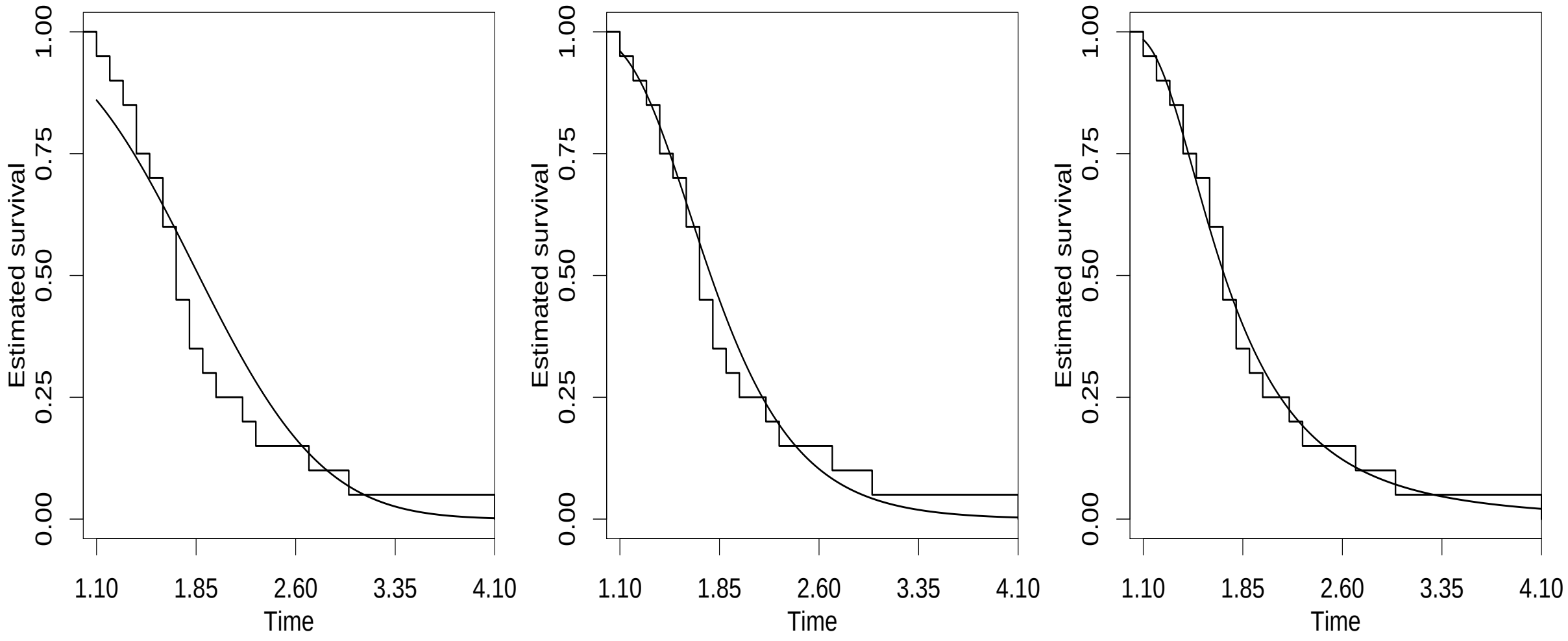

(a) PL distribution (b) GL distribution (c) PIL distribution

Figure 8: Fitted distributions — data set 1.

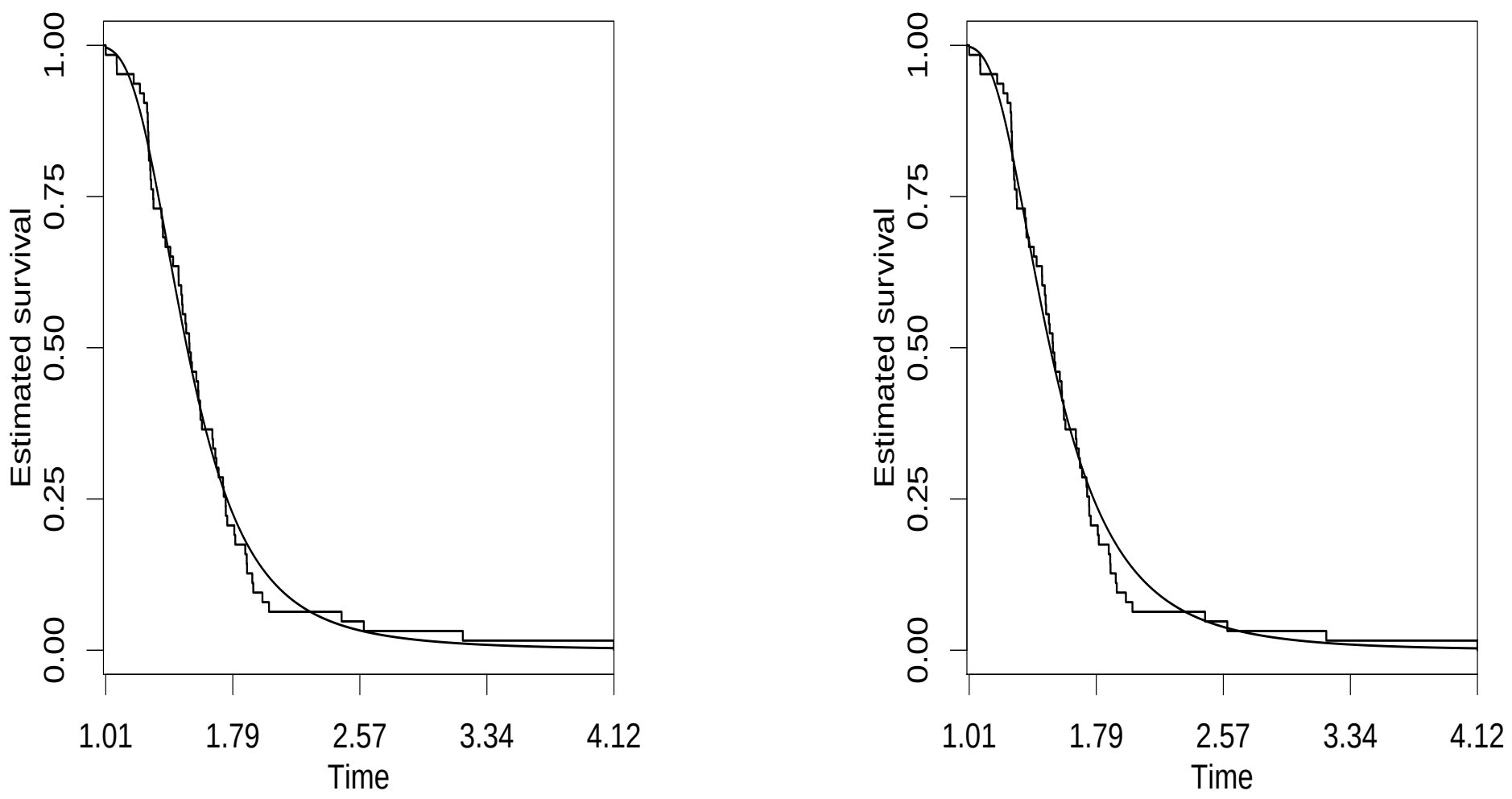

(a) TF distribution

(b) PIL distribution

Figure 9: Fitted distributions — data set 2.

## 8. Conclusion

According to the results obtained by the simulation study, one can observe that the maximum likelihood estimators of the power inverse Lindley distribution have good properties, resulting in small values of bias and mean squared error.

Applied to a real data set and compared with other distributions, it is notable that the power inverse Lindley distribution has a good fit to the data, being a strong competitor to the other distributions tested, as well as a good alternative in modeling survival data.